\documentclass[12pt]{iopart}

\usepackage{ifpdf}
\usepackage[utf8]{inputenc}
\usepackage[english]{babel}
\usepackage{rotating}
\usepackage{graphicx}
\begin{document}
\title
{Incommensurate Magnetic Order in TbTe$_3$}
\author{F. Pfuner$^1$, S.N. Gvasaliya$^{1,2}$, O. Zaharko$^2$, L. Keller$^2$, 
        J. Mesot$^{1,2}$, V. Pomjakushin$^2$, J.-H. Chu $^{3,4}$, 
        I.R. Fisher$^{3,4}$, L. Degiorgi$^1$
        }
\address{$^1$ Laboratory for Solid State Physics, ETH Z\"urich, CH-8093 Z\"urich, Switzerland\\
         $^2$ Laboratory for Neutron Scattering, PSI, CH-5232 Villigen, Switzerland\\ 
         $^3$ Geballe Laboratory for Advanced Materials and Department of Applied 
              Physics, Stanford University, Stanford, California 94305, USA\\
         $^4$ Stanford Institute of Energy and Materials Science, SLAC National 
              Accelerator Laboratory, 2575 Sand Hill Road, Menlo Park 94025, 
              California 94305, USA}
\ead{sgvasali@phys.ethz.ch}
\begin{abstract}
We report a neutron diffraction study of the magnetic phase transitions 
in the charge-density-wave (CDW) TbTe$_3$ compound. We discover that in the 
paramagnetic phase there are strong 2D-like magnetic correlations, 
consistent with the pronounced anisotropy of the chemical structure. 
A long-range incommensurate magnetic order emerges in TbTe$_3$ at 
$T_{mag1}$ = 5.78~K as a result of continuous phase transitions. We 
observe that near the temperature $T_{mag1}$ the magnetic Bragg peaks 
appear around the position (0,0,0.24) (or its rational multiples), that is 
fairly close to the propagation vector $(0,0,0.29)$ associated with the CDW 
phase transition in TbTe$_3$. This suggests that correlations 
leading to the long-range magnetic order in TbTe$_3$ are linked to the 
modulations that occur in the CDW state. 
\end{abstract}
\maketitle
\section{Introduction}
The RTe$_3$ family of compounds (R=rare earth) have recently attracted renewed interest 
as a model system for layered (two-dimensional) charge-density-wave (CDW) materials. 
RTe$_3$ crystallizes in an orthorhombic structure, composed of double layers of planar 
Te- sheets separated by corrugated R-Te layers. The average chemical structure 
can be described using the $Cmcm$ space group~\cite{norling1966} (in this 
setting the lattice parameters $a$ and $c$ are close to 
each other, whereas $b$ is approximately 6 times longer), although in these materials 
an incommensurate lattice modulation with a wave vector $q_c \sim 2/7c^*$ is observed 
below the phase transition into the CDW state~\cite{dimasi1995,malliakas2005}. Such a 
CDW state with modulations along the $c$-axis is present in all RTe$_3$ materials. 
However, the members with the heaviest rare-earth elements exhibit a second 
CDW phase transition at lower temperatures that is characterized by a modulation 
wavevector $q_a \sim 1/3a^*$~\cite{ru2008,fang2007}. The temperatures of these 
two CDW transitions show opposite trends as a function of the R-atom mass. 
Whereas $T_{CDW1}$ decreases towards heavier members of the family ({\it i.e.}, 
going from La to Tm), the $T_{CDW2}$ associated with the $a$- 
axis incommensurability is higher for heavier compounds~\cite{ru2008,ru2008_2}. 
For the specific compound studied in this report, TbTe$_3$, transport and diffraction 
experiments yield $T_{CDW1}$ = 336 K. Although transport measurements do not clearly 
reveal the presence of a second CDW transition, STM measurements performed at 6 K do 
show an ordering pattern that is associated with the wavevector 
$\sim2/3a^*$~\cite{fang2007}. More recent high resolution x-ray diffraction 
measurements confirm the presence of a transverse CDW at low temperatures, with 
wave-vector $\sim 0.683 a^*$, though $T_{CDW2}$ is at present unknown~\cite{banerjee}.

The RTe$_3$ family appears to be a classic CDW system, for which the CDW 
wavevector corresponds to an enhancement of the generalized susceptibility 
$\chi(q)$, which is largely driven by FS 
nesting~\cite{brouet2008,laverock2005,mazin2008}. The enhancement is, however, 
far from being a divergence, suggesting an important role for strong electron-phonon 
coupling ~\cite{mazin2008}. The Fermi surface is composed of states associated with 
the double Te layers, and is therefore similar for all members of the series. 
The principal role of the rare earth ion R appears to be chemical pressure, 
which only modestly affects the resulting CDW wavevectors ~\cite{dimasi1995,ru2008}.

In addition to hosting an incommensurate CDW, for magnetic rare earth ions, RTe3 
also hosts long range antiferromagnetic order at low temperatures. The magnetic 
phase transitions of RTe$_3$ have been studied using 
magnetisation~\cite{ru2008_2,ru2006,iyeiri2003,deguchi2009}, 
calorimetry~\cite{ru2008_2,ru2006,deguchi2009} and electrical 
resistivity~\cite{ru2008_2,ru2006,iyeiri2003,deguchi2009}. Using these 
macroscopic techniques the temperatures of the magnetic transitions have been 
determined and a phase diagram summarizing the temperatures of CDW and 
magnetic phase transition of RTe$_3$ has been established in Ref.~\cite{ru2008_2}. 
However, the nature of the long-range magnetic order in this family of compounds  
has not yet been studied by diffraction techniques, and the effect of the 
incommensurate lattice modulation on the magnetic structure is unknown. To bridge 
this gap we have performed neutron scattering experiments in the representative 
compound TbTe$_3$. The choice of this material was dictated by the following reasons. In 
order to establish possible relationship of the propagation vectors associated to the 
CDW state and of the long-range magnetic order one needs to study the material that has 
reasonably close values between $T_{CDW2}$ and $T_{N}$. Based on this condition, the 
DyTe$_3$ and TbTe$_3$ are both attractive candidates according to their phase 
diagram~\cite{ru2008_2}. DyTe$_3$ undergoes two magnetic phase transitions at 
T$\sim$3.6~K and T$\sim$3.45~K~\cite{ru2008_2}. However, due to high absorption of Dy 
the neutron measurements on DyTe$_3$ are too challenging~\cite{comment1}. 
TbTe$_3$ exhibits three closely located magnetic phase transitions at the 
following temperatures: $T_{mag1}$ = 5.78~K, $T_{mag2}$ = 5.56~K and $T_{mag3}$ =
5.38~K~\cite{ru2008_2}. Here, we present a study of the 
magnetic phase transitions in TbTe$_3$ using powder and single crystal neutron 
diffraction. 

\section{Experimental techniques}
Single crystals of TbTe$_3$ were grown by slow cooling a binary melt, as 
described elsewhere~\cite{ru2006}. As the material is rather air-sensitive, the 
TbTe$_3$ powder was obtained by crushing the single crystals in He-atmosphere. The 
powder was immediately sealed in a cylindric vanadium can of 6~mm in diameter 
used for the measurements. 

Powder neutron diffraction experiments were performed at the DMC cold-neutron 
diffractometer~\cite{DMC}(neutron spallation source SINQ~\cite{SINQ}, PSI) in the 
temperature range 1.5 -- 60~K. Most of the data were collected with 
neutron wavelengths of $\lambda=2.46$~\AA , giving access to the wave vectors range 
0.1--3.4~\AA $^{-1}$. Typical exposure times were $\sim12$ hours. 

The experiments on the single crystals of TbTe$_3$ with size of roughly 
$\sim 7 \times 8 \times 2.5$ mm$^3$ were performed at the TriCS 
single crystal diffractometer~\cite{TriCS}. A pyrolytic graphite (PG) 
monochromator was used with neutron wavelength $\lambda=2.32$ \AA. A 120 mm long 
PG filter was installed in the beam in order to ensure full suppression of 
higher-order neutrons. The temperature range of interest for these measurements is 
$T\sim2 - 8$~K, because the three magnetic phase transitions are close to each 
other in the range $5-6$~K. This interval was scanned with steps of 
$\sim 0.05$~K. To ensure the required accuracy in temperature an "orange"-type He flow cryostat 
was used. One of the samples was aligned in the [H,K,0] plane and a second sample 
was aligned in the [0,K,L]. The normal beam geometry was employed, tilting the 
detector in the range of -10...10 deg. This allowed to access reflections with 
L=$\pm$1 and H=$\pm$1, respectively.

\section{Powder Diffraction Results}
\begin{figure}
\begin{center}
\includegraphics[scale=0.5]{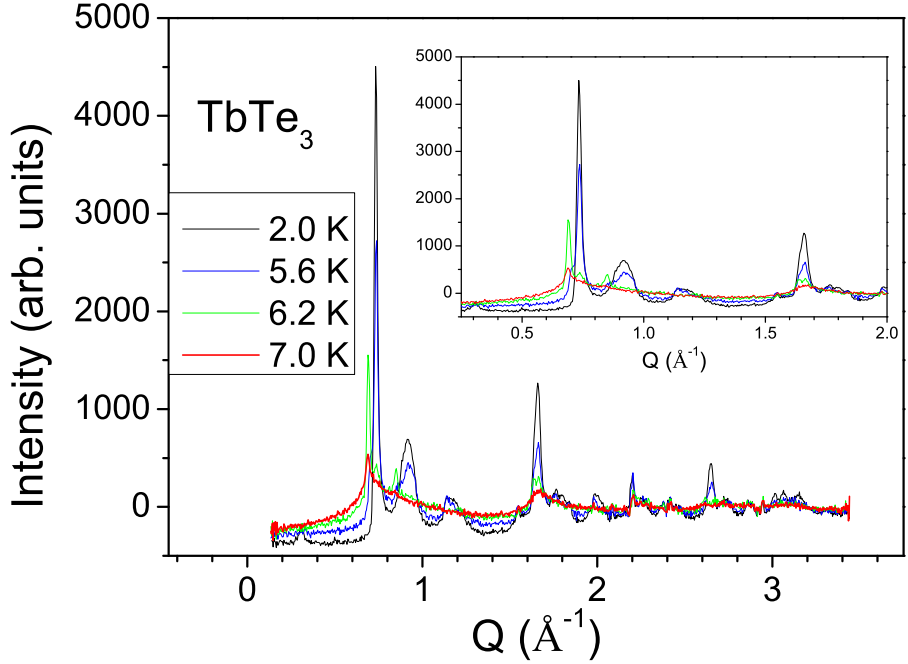}
\caption{Temperature dependence of the difference-patterns with respect to the data 
         at the reference temperature of 60~K. The inset emphasizes the changes 
         of the diffraction pattern in the range of wavevectors 0.2 - 2\AA {\ }.
}
\label{differences}
\end{center}
\end{figure}
\begin{figure}
\begin{center}
\includegraphics[scale=0.5]{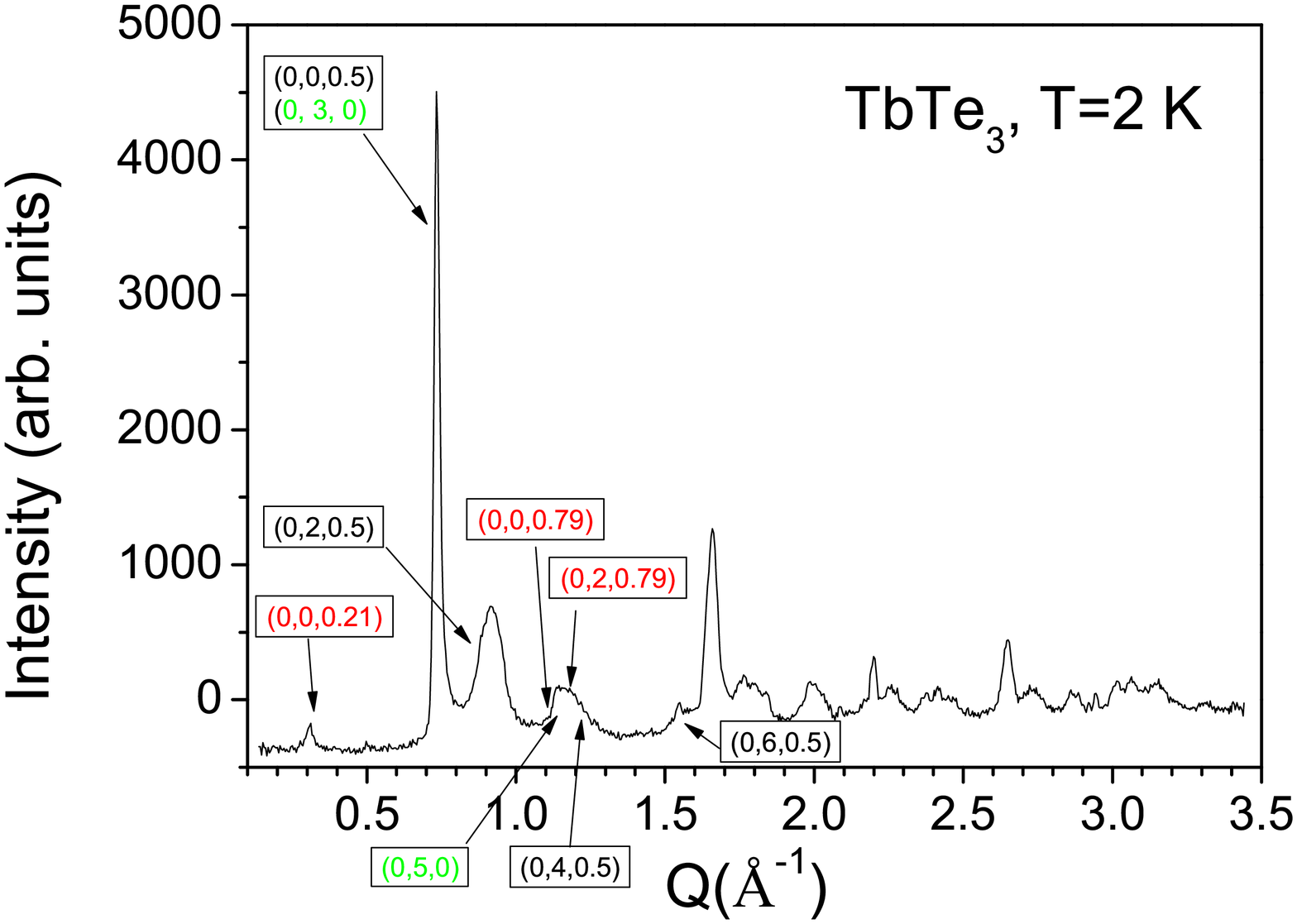}
\caption{Indexed magnetic reflections for TbTe$_3$ at 2 K in a powder diffraction pattern 
         (the contribution from the chemical structure is eliminated by subtracting the 
         powder spectrum taken at T=60 K). The 
         indexed peaks were identified by single-crystal neutron diffraction experiments. 
         Labels in the brackets are the ($H,K,L$) indices of the 
         peaks given in terms of the average $Cmcm$ structure. The peaks that are 
         associated with the magnetic propagation vector q$^{mag}_1=(0,0,0.21)$ 
         are labeled in red, those associated with the magnetic propagation vector 
         q$^{mag}_2=(0,0,0.5)$ are shown in black, and finally, with  
         q$^{mag}_3=(0,1,0)$ - in green. Note that the 
         Bragg peaks that are labeled for simplicity as ($0,0,0.21$), in fact have 
         a non-zero component of modulation along the ($<0,K,0>$) direction and more 
         precisely should be described as ($0,0.01,0.21$). See text and 
         Fig.~\ref{fits_000p21} for details. 
}
\label{Fig_DMC_diff}
\end{center}
\end{figure}
\noindent The data at 60 K was taken as a reference for the scattering in 
the paramagnetic phase of TbTe$_3$. The average chemical structure of TbTe$_3$ was refined 
with Fullprof~\cite{fullprofref} within the $Cmcm$ space group following a structural model 
of the RTe$_3$-type materials 
developed in Ref.~\cite{malliakas_2006}. This analysis indicates the presence of several preferred orientations in the sample, impeding high-quality refinements 
of both chemical and 
magnetic structures. Nevertheless, the lattice 
parameters obtained from the data $a=4.3050\pm0.00015$\AA{\ }, 
$b= 25.33757\pm0.0012$\AA{\ }, and 
$c= 4.28024\pm0.00012$\AA{\ } are in agreement with those published 
earlier~\cite{malliakas_2006}. In what follows we concentrate on the differences of 
the data taken at temperatures lower than 60~K. In this experimental setup the 
low-temperature changes in the chemical structure of TbTe$_3$ are not resolved. 

Figure~\ref{differences} shows difference-patterns at selected temperatures, 
obtained by eliminating the contribution 
from the chemical structure with the subtraction of the diffraction pattern at 
60 K. As expected from the macroscopic properties of TbTe$_3$, magnetic Bragg peaks 
appear in the diffraction patterns collected below $T_{mag1}$= 5.78~K. 
Furthermore, relatively sharp asymmetric peaks appear in the diffraction 
patterns even at temperatures slightly higher then $T_{mag1}$ (e.g., 
the patterns taken at 6.2~K and 7~K). As shown in Fig.~\ref{differences}, these 
unusual peaks have an extended shoulder towards higher wavevectors. 
Such scattering profiles from powders are typical for systems 
with pronounced quasi-two-dimensional (quasi-2D) properties, as observed {\it e.g.} 
in graphite~\cite{warren_1941} or in 2D magnetic 
materials~\cite{roessli_1993}. Therefore, one may expect that in 
a relatively broad temperature range above $T_{mag1}$ TbTe$_3$ exhibits 
quasi-2D behavior. The positions of the magnetic Bragg peaks 
are temperature-dependent and may be clearly distinct in different magnetic phases. 
This suggests that the magnetic phases occurring in TbTe$_3$ are modulated and 
the propagation vector is locked-in at incommensurate position only below 
$T_{mag3}$. In addition, the asymmetric peaks above $T_{mag1}$ are located at a 
different wavevector as compared to magnetic Bragg peaks observed just 
below $T_{mag1}$. This is best seen for the scattering profiles 
shown in Fig.~\ref{differences} in the vicinity of 
${\bf Q} \sim 0.7^{-1}$\AA{\ }. 

Figure~\ref{Fig_DMC_diff} shows the magnetic diffraction pattern measured from 
TbTe$_3$ at 2~K. Some of the magnetic peaks could be indexed in terms of the 
average chemical cell and are labeled in Fig.~\ref{Fig_DMC_diff} (due to the complexity of 
the pattern the use of single-crystal neutron diffraction data was essential for this step).
Clearly, most of the magnetic Bragg peaks are at positions incommensurate with the 
average chemical structure of the material. It may be speculated that the 
broad magnetic peaks observed in the vicinity 
of $Q \simeq 0.9 $\AA{\ } and $Q \simeq 1.2$\AA{\ } are composed by many 
magnetic Bragg peaks which are located close to each other in wave-vector space, 
due to the rather large value of the lattice parameter $b$. We note here that among 
the magnetic Bragg peaks we could not identify any with non-zero value of $H$.

\section{Single-Crystal Diffraction Results}
\begin{figure}
\begin{center}
\includegraphics[scale=0.4]{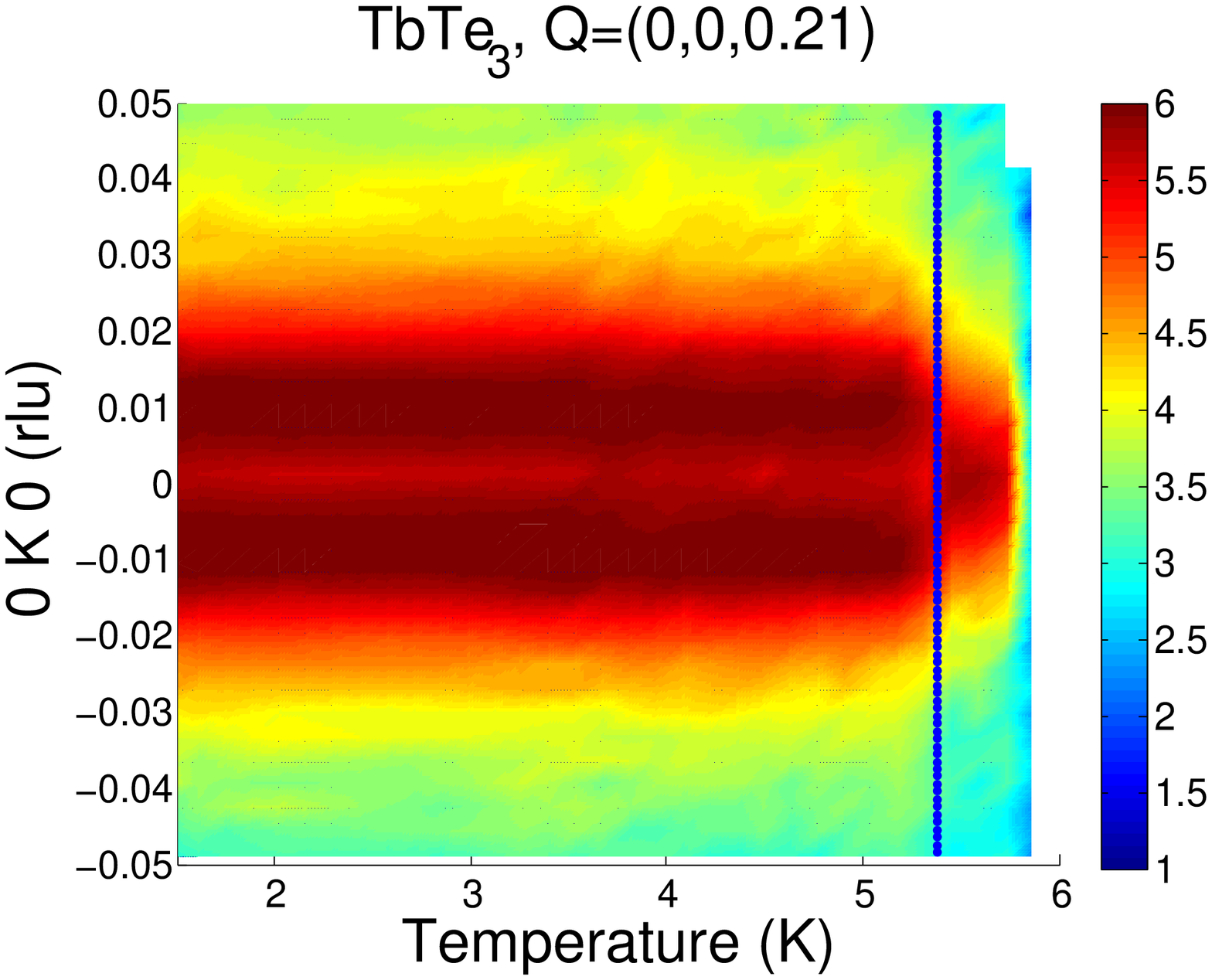}
\includegraphics[scale=0.4]{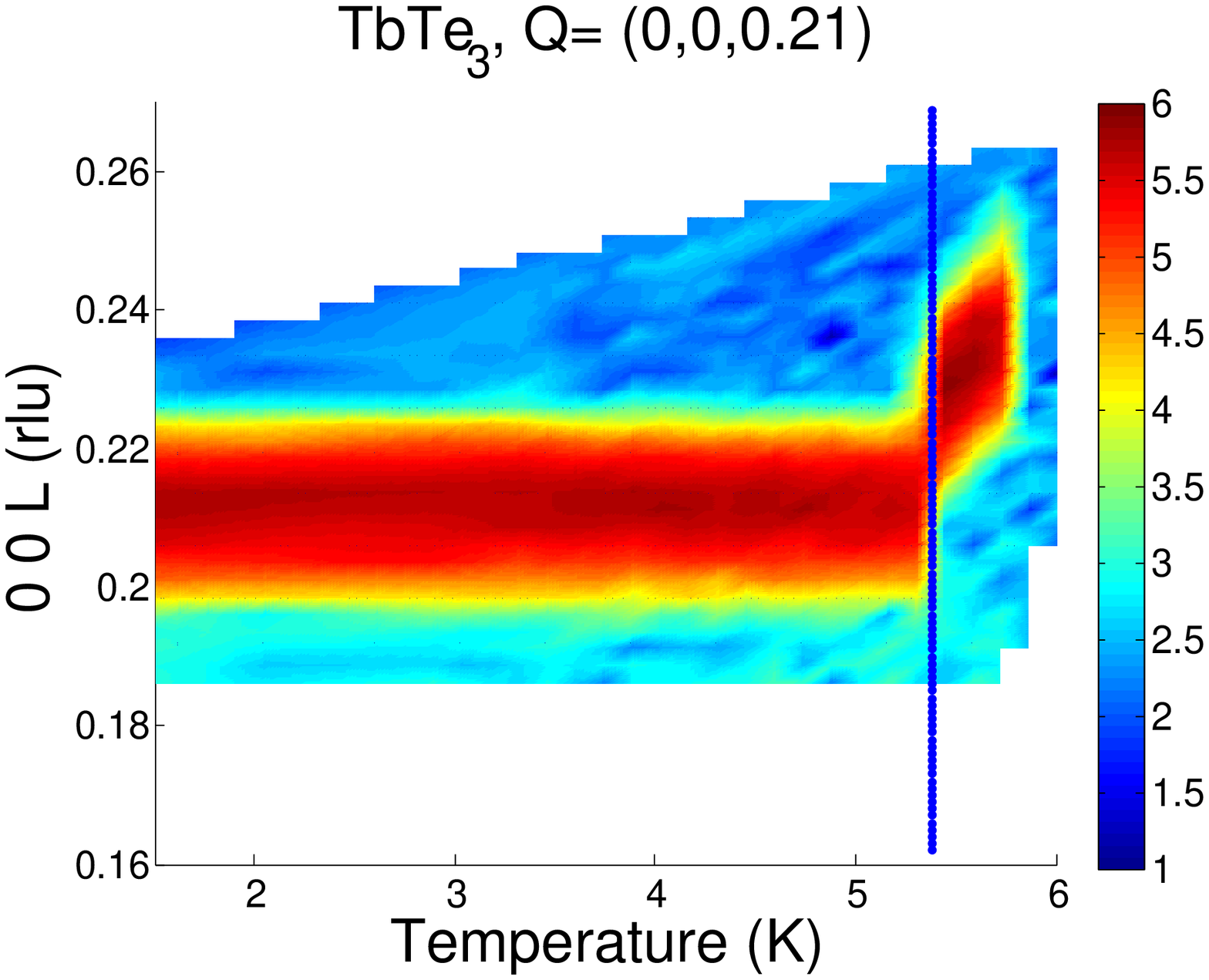}
\caption{The temperature evolution of the intensity distribution taken in 
         the vicinity of the (0,0,0.21) position of TbTe$_3$. Vertical solid lines on
         both plots denote $T_{mag3}$ = 5.38~K. Note that below 
         $T_{mag3}$ the reflection is incommensurate in both 
         $<0,K,0>$ and $<0,0,L>$ directions. 
}
\label{maps_000p21}
\end{center}
\end{figure}

\noindent Taking into account the complexity of the powder diffraction patterns 
observed in magnetically ordered phases of TbTe$_3$, there are considerable 
difficulties not only in developing a proper model of the arrangement of the 
Tb magnetic moments, but even in determining the propagation 
vectors in the magnetically ordered phases. In a 
trial to overcome these difficulties a series of single-crystal experiments 
were performed at the base temperature $T$=2~K, below $T_{mag3}$. Those 
reflections, which were successfully indexed at $T$=2~K, are labeled in 
Fig.~\ref{Fig_DMC_diff}. There was no attempt to index the peaks observed 
in powder diffraction patterns at higher temperatures. Instead, the 
temperature evolution of those peaks that were identified at $T$= 2 K 
was traced up to $T\sim7$~K. 
\begin{figure}
\begin{center}
\includegraphics[scale=0.45]{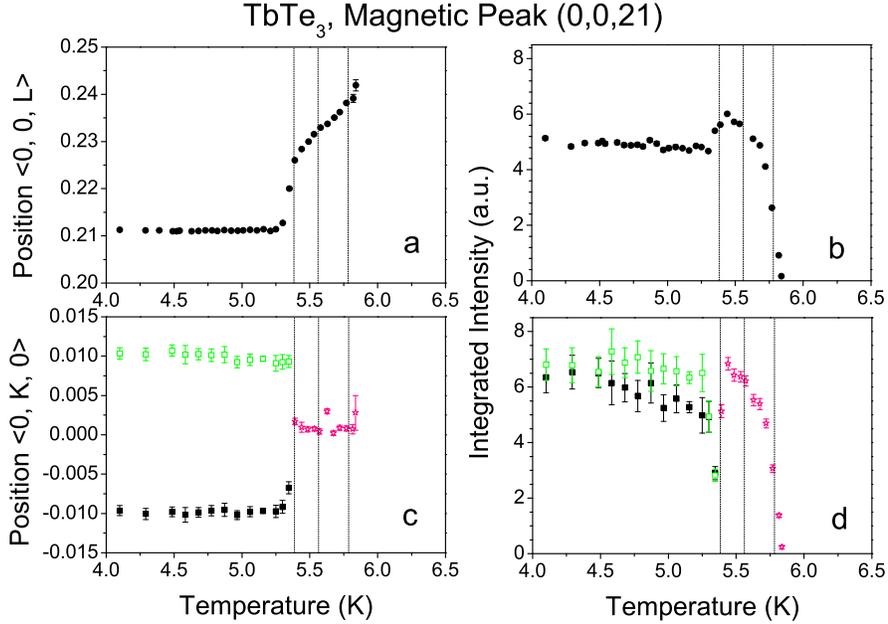}
\caption{The temperature evolution of the propagation vectors (left column) 
         and of the intensity (right column) measured around the position 
         (0,0,21). Green symbols denote the position and the intensity of the 
         (0,$0+\delta$,0.21) satellite, whereas black squares stand for the 
         (0,$0-\delta$,0.21). Vertical dashed lines denote the temperatures 
          of the magnetic phase transitions.
}
\label{fits_000p21}
\end{center}
\end{figure}

Figure~\ref{maps_000p21} shows false-color maps of neutron intensities 
taken in a close vicinity of the ($0,0,0.21$) incommensurate magnetic 
reflection in the temperature range $1.5 - 6$~K. At the base temperature this 
reflection has incommensurate components both along the $<0,K,0>$ 
and $<0,0,L>$ directions. However, the incommensurate component along 
the $<0,K,0>$ direction appears to be small and diminishes above $T_{mag3}$ 
in a step-like way. 
Above $T_{mag3}$ the incommensurate components along the $<0,0,L>$ 
direction is also temperature-dependent. For a more quantitative analysis the scans 
for each temperature are fitted assuming a Gaussian shape for a Bragg reflection.
The resulting temperature dependence of the position and intensity of the peak ($0,0,0.21$) 
are shown in Fig.~\ref{fits_000p21}. The modulation vector along the $<0,0,L>$ 
direction stays constant within the precision of the measurements below $T_{mag3}$.
Above $T_{mag3}$ the propagation vector increases smoothly and reaches the value 
of $(0, 0, 0.242\pm0.001)$ at $T_{mag1}$. The modulation vector along the $<0,K,0>$ 
direction is also constant $(0, 0.01\pm0.001, 0)$ below $T_{mag3}$, while above 
this temperature the component of the modulation vanishes. The intensity measured 
around the $(0,0,0.21)$ position also appears to 
be nearly temperature-independent below $T_{mag3}$. Above this temperature however, 
the overall intensity observed in the vicinity of the $(0,0,0.21)$ position 
drops by a factor of 2, as can be deduced from Fig.~\ref{fits_000p21}b-d. Further the 
intensity gradually decreases and above $T_{mag1}$ we could not detect any scattering 
above the background level around that position. In the whole studied temperature 
range the width of the peaks is limited by resolution. Gradual decrease of the 
Bragg intensity suggest a continuous character of the phase transition that the crystal 
undergoes at $T_{mag1}$, whereas an abrupt change in magnetic intensity at $T_{mag3}$ 
points to a first order phase transition. 

\begin{figure}
\begin{center}
\includegraphics[scale=0.26]{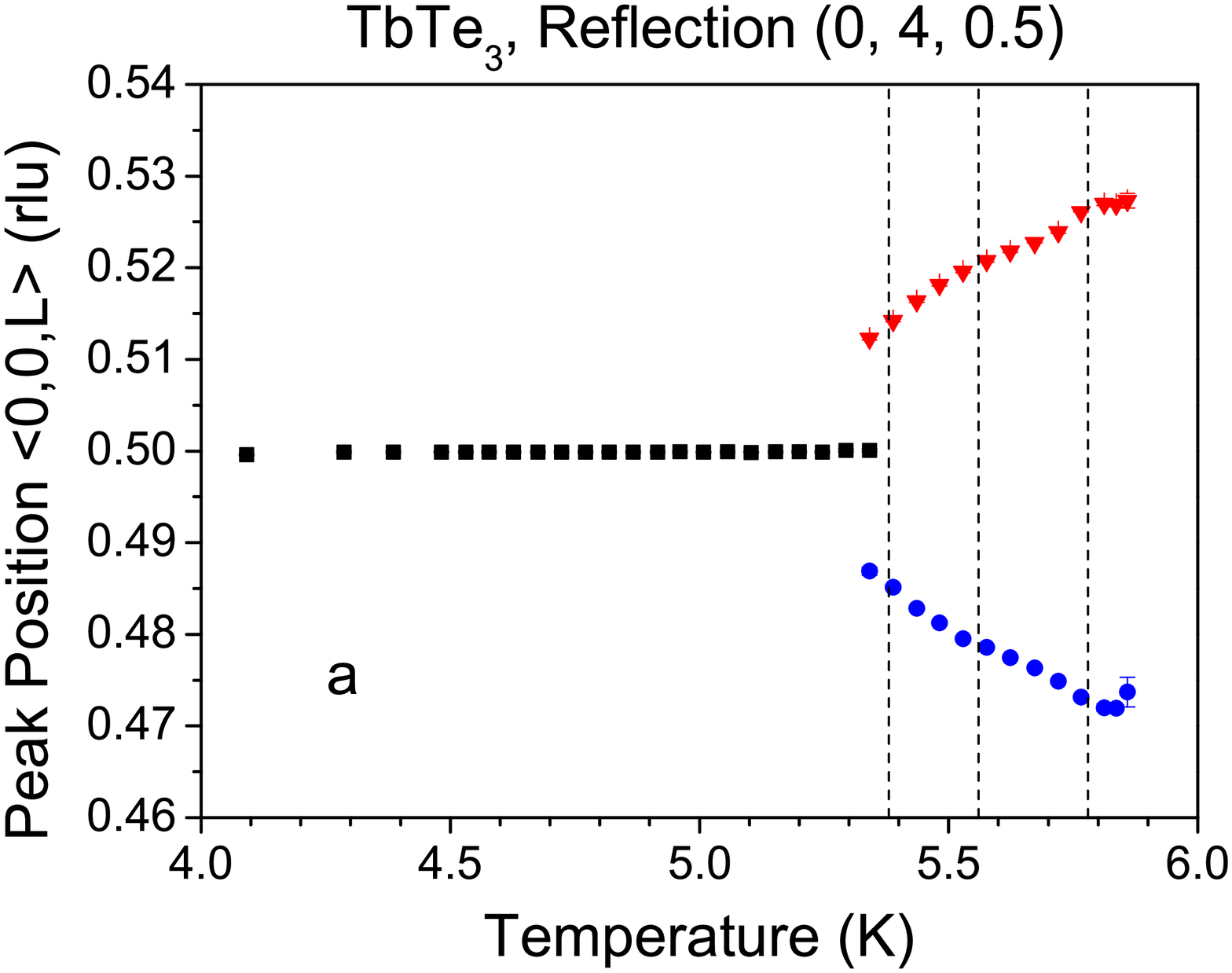}
\includegraphics[scale=0.26]{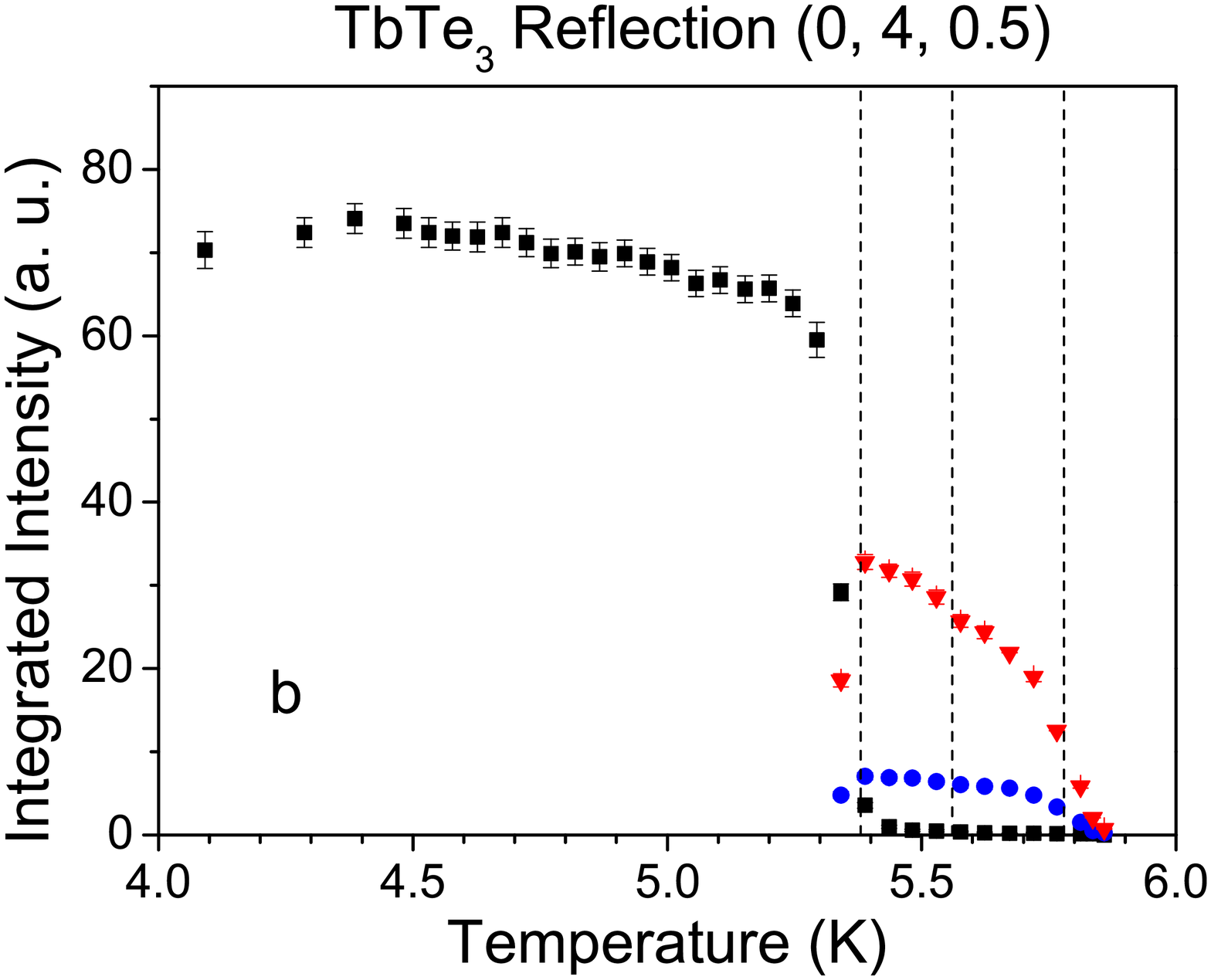}
\caption{The temperature evolution of the intensity distribution taken in 
         the vicinity of the (0,4,0.5) position of TbTe$_3$. Red triangles 
         show the position and the intensity of (0,0,$0.5+\delta$) satellite, 
         while blue circles denote the position and the intensity of 
         (0,0,$0.5-\delta$) satellite. Vertical dashed lines denote
         the temperatures of the magnetic phase transitions.
}
\label{fits_040p5}
\end{center}
\end{figure}

The magnetic Bragg peaks, characterized by 
the propagation vector (0,0,0.5), exhibit a qualitatively different behavior. 
Figure~\ref{fits_040p5} shows the temperature 
dependence of the $(0,4,0.5)$ Bragg peak in the temperature range $4 - 6$~K 
as a representative example. Whereas below $T_{mag3}$ the propagation vector 
is (0,0,0.5), above this 
temperature the position of the magnetic Bragg peaks becomes incommensurate in a step-like 
fashion as (0,0,0.5$\pm\delta$) with $\delta$ displaying a temperature dependence. 
The intensity of the $(0,4,0.5)$ magnetic peak decreases abruptly at $T_{mag3}$ and 
smoothly diminishes towards $T_{mag1}$. Weak magnetic peaks around the $(0,4,0.5)$ 
position are still observed in paramagnetic phase up to $\sim6$~K. 
 
\section{Discussion and Conclusions}
We have performed a first study of magnetic phase transitions in TbTe$_3$ 
by powder and single-crystal neutron diffraction. We find that in the paramagnetic 
phase, slightly above $T_{mag1}$ = 5.78~K, there are pronounced 
2D-like magnetic correlations. At $T_{mag1}$ long-range magnetic order emerges 
as a result of a continuous phase transition. 
In all three magnetically ordered phases, incommensurate modulations are present. 
This observation is at variance with the typical behavior of unmodulated 
rare-earth intermetallic compounds 
(see {\it e.g.}~\cite{gignoux1993,gignoux1994}).  Incommensurate magnetic 
structures often appear in these materials just below the phase transition into 
the magnetically ordered state. However, in most cases magnetoelastic coupling 
and crystal-field effects typically result in a lock-in transition to a commensurate 
magnetic structure. The behavior found in TbTe$_3$ only partially resembles these 
general expectations. As shown in Fig.~\ref{fits_040p5} the magnetic propagation 
vector of the $(0,0,0.5\pm\delta)$-type observed just below $T_{mag1}$ turns out to 
be locked-in below $T_{mag3}$ into the simple antiferromagnetic position 
$(0,0,0.5)$. However, the magnetic Bragg peaks observed below $T_{mag1}$ near the 
position (0,0,0.24), stabilize at low temperature at the incommensurate position 
(0,$0+\delta$,0.21). This effect is possibly due to the incommensurate lattice 
modulation present in TbTe$_3$ in the CDW state. 
 
The propagation vectors of the magnetic structures, which we succeeded to 
identify in TbTe$_3$, have components either along the $<0,K,0>$ or $<0,0,L>$ directions, 
or a linear combination of both. We could not assign any of the magnetic peaks, 
observed in powder diffraction pattern, as a Bragg reflection with non-zero 
component $<H,0,0>$ of the propagation vector. Therefore, our results do not 
point towards a link of the propagation vector associated with the second CDW phase 
transition and of the magnetic order in TbTe$_3$. Nonetheless, we discover that near 
the temperature $T_{mag1}$ the magnetic Bragg peaks appear around the positions 
(0,0,0.24) (or its rational multiples). This value is fairly close to the propagation 
vector $(0,0,0.29)$~\cite{ru2008} associated with the high-temperature CDW phase 
transition, raising the possibility that correlations leading to the 
long-range magnetic order in TbTe$_3$ might be linked to the modulated chemical 
structure in the CDW state. These results stimulate further work to refine the 
magnetic structure in this and related magnetic RTe$_3$ compounds, with the ultimate goal 
of establishing how the CDW and magnetic order interact and coexist.  

\section{Acknowledgments}
This work is based on the experiments performed at the Paul Scherrer Institut, 
Switzerland and has been supported by the Swiss National Foundation for the 
Scientific Research within the NCCR MaNEP pool. Work at Stanford University 
was supported by the Department of Energy, Office of Basic Energy Sciences under 
contract DE-AC02-76SF00515. 

\section*{References}
\begin{thebibliography}{99}
\bibitem{norling1966} Norling B K and Steinfink H 1966 Inorg. Chem. {\bf5}, 1488 
\bibitem{dimasi1995} DiMasi E, Aronson M C, Mansfield J F, Foran B and Lee S 1995  
Phys. Rev. B {\bf52}, 14516
\bibitem{malliakas2005} Malliakas C, Billinge S J L, Kim H J and Kanatzidis M G 2005 
J. Am. Chem. Soc. {\bf 127}, 6510 
\bibitem{ru2008} Ru N, Condron C L, Margulis G Y, Shin K Y, Laverock J, Dugdale S B, 
Toney M F and Fisher I R 2008 Phys. Rev. B {\bf 77}, 035114 
\bibitem{fang2007} Fang A, Ru N, Fisher I R and Kapitulnik A 2007 Phys. Rev. Lett. 
{\bf99}, 046401 
\bibitem{ru2008_2} Ru N, Chu J H and Fisher I R 2008 Phys. Rev. B {\bf78}, 012410 
\bibitem{banerjee} A. Banerjee and T. Rosenbaum, private communication
\bibitem{brouet2008} Brouet V, Yang W L, Zhou X J, Hussain Z, Moore R G, He R, 
Lu D H, Shen Z X, Laverock J, Dugdale S B, Ru N, and Fisher I R 2008 Phys. Rev. 
B {\bf77}, 235104 
\bibitem{laverock2005} Laverock J, Dugdale S B, Major Zs, Alam M A, Ru N, 
Fisher I R, Santi G and Bruno E 2005 Phys. Rev. B {\bf 71}, 085114 
\bibitem{mazin2008} Johannes M D and Mazin I I 2008 Phys. Rev. B {\bf 77}, 165135 
\bibitem{ru2006} Ru N and Fisher I R 2006 Phys. Rev. B {\bf73}, 033101 
\bibitem{iyeiri2003} Yuji Iyeiri, Teppei Okumura, Chishiro Michioka and Kazuya Suzuki 2003 
Phys. Rev. B {\bf67}, 144417 
\bibitem{deguchi2009} Deguchi K, Okada T, Chen G F, Ban D, Aso N and Sato N K 2009 
Journal of Physics Conference Series {\bf150}, 042023 
\bibitem{comment1} In fact we gave a trial and performed a neutron powder diffraction 
experiment on DyTe$_3$ packed into a double wall vanadium container. However, with the 
exception of two most intense peaks, the signal to noise ratio was not favorable for 
reliable conclusions. Therefore, we do not report the neutron 
scattering results obtained for DyTe$_3$ and concentrate here on TbTe$_3$. 
\bibitem{DMC} Fischer P, Keller L, Schefer J and Kohlbrecher J 2000 Neutron News {\bf11}, 19 
\bibitem{SINQ} Blau B et al., 2009 Neutron News {\bf20}, 5 
\bibitem{TriCS} Schefer J, K\"onnecke M, Murasik A, Czopnik A, Str\"assle Th, Keller P and 
Schlump N 2000 Physica B {\bf 276-278}, 168 
\bibitem{fullprofref} Rodr\'iguez-Carvajal J 1993 Physica B {\bf 192}, 55  
\bibitem{malliakas_2006} Malliakas C and Kanatzidis M 2006 J. Am. Chem. Soc. {\bf 128}, 12612 
\bibitem{warren_1941} Warren B E 1941 Phys. Rev. {\bf 59}, 693
\bibitem{roessli_1993} Roessli B, Fischer P, Zolliker M, Allenspach P, Mesot J, 
Staub U, Furrer A, Kaldis E, Bucher B, Karpinski J, Jilek E and Mutka H 
1993 Z. Phys. B {\bf 91}, 149 
\bibitem{gignoux1993} Gignoux D and Schmitt D 1993 Phys. Rev. B {\bf 48}, 12682 
\bibitem{gignoux1994} Gignoux D and Schmitt D 1994 JMMM {\bf 129}, 53
\end{thebibliography}
\end{document}